\documentclass[reprint,aps,pra,10pt,superscriptaddress,amsmath,amssymb,floatfix]{revtex4-2}

\usepackage{amsmath,amssymb,bm}
\usepackage{mathtools}
\usepackage{esvect}
\usepackage{booktabs}
\usepackage{multirow}
\usepackage{array}
\usepackage{upgreek}
\usepackage[svgnames]{xcolor}
\usepackage[caption=false]{subfig}

\begin{document}

\title{Isotope-shift analysis with the $4f^{14}6s^{2}~^1S_0- 4f^{13}5d6s^{2}(J=2)$ transition in ytterbium}
\author{Akio Kawasaki}
\email{akio.kawasaki@aist.go.jp}
\affiliation{National Metrology Institute of Japan (NMIJ), National Institute of Advanced Industrial Science and Technology (AIST), 1-1-1 Umezono, Tsukuba, Ibaraki 305-8563, Japan}

\author{Takumi Kobayashi}
\affiliation{National Metrology Institute of Japan (NMIJ), National Institute of Advanced Industrial Science and Technology (AIST), 1-1-1 Umezono, Tsukuba, Ibaraki 305-8563, Japan}

\author{Akiko Nishiyama}
\affiliation{National Metrology Institute of Japan (NMIJ), National Institute of Advanced Industrial Science and Technology (AIST), 1-1-1 Umezono, Tsukuba, Ibaraki 305-8563, Japan}

\author{Takehiko Tanabe}
\affiliation{National Metrology Institute of Japan (NMIJ), National Institute of Advanced Industrial Science and Technology (AIST), 1-1-1 Umezono, Tsukuba, Ibaraki 305-8563, Japan}

\author{Masami Yasuda}
\affiliation{National Metrology Institute of Japan (NMIJ), National Institute of Advanced Industrial Science and Technology (AIST), 1-1-1 Umezono, Tsukuba, Ibaraki 305-8563, Japan}


\begin{abstract}
Measurements of isotope shifts have recently been attracting considerable attention due to their potentials in searching for new forces. We report on the isotope shifts of the $4f^{14}6s^{2}~^1S_0- 4f^{13}5d6s^{2}(J=2)$ transition at 431 nm in Yb, based on absolute frequency measurements with an accuracy of $\sim10$ kHz. With these data, the hyperfine constants for $^{173}$Yb are determined. To analyze these data further, electronic structure of ytterbium is theoretically calculated. The nuclear charge radii are estimated together with some previously reported isotope-shift data for other transitions. An analysis of the King plot for the $4f^{14}6s^{2}~^1S_0- 4f^{13}5d6s^{2}(J=2)$ transition shows a good consistency with other transitions, resulting in a constraint on the existence of new bosons mediating the force between an electron and a neutron. The analysis motivates further precision measurements on isotope shifts of the narrow-linewidth transitions in ytterbium, not only for the even-mass isotopes but also for the odd-mass isotopes. 
\end{abstract}

\maketitle

\section{Introduction}
Isotopes, which differ in the number of neutrons $N$ in the nucleus within the same element, exhibit slightly varying physical properties. Such properties include macroscopic quantities, e.g., phase transition temperatures, reaction rates of chemical reactions, and diffusion constants, as well as nuclear properties specific to nuclei. The nuclear properties manifest themselves particularly well in the difference in the resonant frequency $\nu$ for a transition $\alpha$, which is called an isotope shift. Typically ranging from 100 MHz to 1 GHz, the isotope shift for an optical transition of $\sim100$~THz is easily observed with laser spectroscopy. 

Isotope shifts are conventionally attributed primarily to two sources. One is the mass difference between nuclei changing the reduced mass and correlation of electrons, which is called the mass shift. The other is the variation of the nuclear charge radius perturbing the Coulomb potential inside the nucleus, referred to as the field shift. Precise determination of isotope shifts has served as a good probe of nuclear structure \cite{AtomDataNuclData.99.69,NuclChargeRadii}. 

In addition to these two main factors, recent developments in analysis of isotope shifts incorporates higher order terms as shown in the following equation \cite{PhysRevLett.128.163201}:
\begin{eqnarray}\label{eqnIS}
\nu_{\alpha}^{AA'}=&F_\alpha \delta\langle r^2\rangle^{AA'}+K_\alpha \mu ^{AA'} + G_\alpha^{(4)}\delta\langle r^4\rangle^{AA'} \nonumber \\
&+ G_\alpha^{(2)}
[(\delta\langle r^2 \rangle)^2]^{AA'}
+ v_{n e} D_\alpha  N^{AA'}
\end{eqnarray}
Here, the superscripted mass number $A$ specifies the isotope, and $X^{AA'} \equiv X^A-X^{A'}$. The subscripted $\alpha$ denotes the transition. Note that transitions are labeled with their wavelengths (three-digit number in the unit of nm) in this paper. All transitions of ytterbium (Yb) appearing in this paper are summarized in Table \ref{tableTransitions}. Coefficients $F_{\alpha}$ and $K_{\alpha}$ characterize the field shift and mass shift, respectively. $\langle r^n\rangle$ is the $n$th moment of the nuclear charge, and $\delta \langle r^n\rangle^{AA'}$ is its difference between two isotopes. $\mu ^{A} =1/m^A$ is the inverse mass. The third and fourth terms are the fourth-moment shift and the quadratic field shift, respectively, where $[(\delta\langle r^2 \rangle)^2]^{AA'}=( \delta\langle r^2\rangle^{AA''} )^2 - ( \delta\langle r^2\rangle^{A'A''} )^2$ with $A''$ being a reference isotope. The fifth term shows the effect of a Yukawa potential $V_{ne}=\hbar c v_{ne}\exp(-m_{\phi}cr/\hbar)/r$ on the electrons besides the Coulomb potential. $v_{n e}=(-1)^{s+1}y_n y_e /(4\pi\hbar c)$ characterizes the strength of a hypothetical new force between a neutron and an electron, where $s$ is the spin of the boson mediating this force, $y_{n,e}$ are coupling constants to a neutron and an electron, and $m_{\phi}$ is the mass of the boson. 

The field-shift term can be eliminated by a set of isotope shifts for another transition $\beta$. The resulting equation is
\begin{eqnarray}\label{EqKing}
\frac{\nu_{\beta}^{AA'}}{\mu^{AA'}}&=&\frac{F_\beta}{F_\alpha} \frac{\nu_{\alpha}^{AA'}}{\mu^{AA'}}+K_{\beta, \alpha} +\frac{\delta\langle r^4\rangle^{AA'}}{\mu^{AA'}} G_{\beta, \alpha}^{(4)}\nonumber \\
&&
+\frac{[(\delta\langle r^2 \rangle)^2]^{AA'}}{\mu^{AA'}} G_{\beta, \alpha}^{(2)}+\frac{v_{n e} N^{AA'}}{\mu^{AA'}} D_{\beta, \alpha}, \label{eqnKing}
\end{eqnarray}
where $X_{\beta, \alpha}=X_{\beta}-X_{\alpha} F_\beta/F_\alpha $ for $X \in\{K, D, G^{(4)},G^{(4)}\}$. The plot between $\nu_{\beta}^{AA'}/\mu^{AA'}$ and $\nu_{\alpha}^{AA'}/\mu^{AA'}$ is called the King plot and is approximately linear \cite{KingIS}, as the first two terms in the right-hand side of Eq. (\ref{EqKing}) are major. The breakdown of the linearity is proposed to be used for searches for the new bosons \cite{PhysRevLett.120.091801,EurPhysJC.77.896,PhysRevA.97.032510}, and some advanced analysis techniques are also reported to reduce the effects with large uncertainties \cite{PhysRevResearch.2.043444,PhysRevLett.125.123002}, along with assessments of which higher order effects are significant \cite{PhysRevA.103.L030801,PTEP.2020.103B02,PhysRevC.101.021301,PhysRevC.102.024326,PhysRevA.104.L020802}. For heavy atoms, normalizing both sides with $\nu_{\alpha}^{AA'}/\mu^{AA'}$ suppresses the influence of the uncertainty in mass \cite{PhysRevLett.125.123002}:
\begin{eqnarray}
\frac{\nu_{\beta}^{AA'}}{\nu_{\alpha}^{AA'}}&=&\frac{F_\beta}{F_\alpha}+K_{\beta, \alpha} \frac{\mu^{AA'}}{\nu_{\alpha}^{AA'}}+\frac{\delta\langle r^4\rangle^{AA'}}{\nu_{\alpha}^{AA'}} G_{\beta, \alpha}^{(4)}\nonumber \\
&&+\frac{[(\delta\langle r^2 \rangle)^2]^{AA'}}{\nu_{\alpha}^{AA'}} G_{\beta, \alpha}^{(2)}+\frac{v_{n e} N^{AA'}}{\nu_{\alpha}^{AA'}} D_{\beta, \alpha} \label{eqnNormKing}
\end{eqnarray}

Among various atoms\cite{PhysRevLett.125.123003,PhysRevA.103.L040801,PhysRevLett.131.161803,2311.17337,PhysRevLett.125.123002,PhysRevLett.128.163201,PhysRevLett.128.073001,PhysRevX.12.021033}, Yb is one of the most extensively studied atomic species on the isotope-shift measurements aimed at searching for new bosons, with the nonlinearity of the King plot \cite{PhysRevLett.125.123002,PhysRevLett.128.163201,PhysRevLett.128.073001,PhysRevX.12.021033}. The isotope shifts for the 436-nm transitions in Yb$^+$ \cite{PhysRevLett.125.123002} and the 361-nm transition in Yb \cite{PhysRevLett.128.073001} are measured with at best 300-Hz accuracy. The 578-nm transition in Yb \cite{PhysRevX.12.021033} and the 411- and 467-nm transitions in Yb$^+$ \cite{2403.07792} have $\lesssim 10$ Hz accuracy measurements. However, these analyses so far only cover even-mass isotopes, because extra nonlinearity potentially induced by the nuclear spins in odd-mass isotopes can become a background in the search for new bosons. In fact, except for the 578-nm transition in Yb, no precise isotope-shift measurements are available for odd-mass isotopes; only broad-linewidth transitions utilized for laser cooling have the isotope-shift data for the odd-mass isotopes. Particularly, regarding a newly observed narrow-linewidth transition at 431 nm \cite{PhysRevA.107.L060801,PhysRevLett.130.153402}, the absolute frequency is measured for only a single odd-mass isotope \cite{PhysRevA.107.L060801}, and isotope shifts are measured for only even-mass isotopes \cite{PhysRevLett.130.153402}.

\begin{table}[t]
	\caption{Transitions in Yb and Yb$^+$. The numbers listed in $\lambda$ are the wavelengths of the transitions in the unit of nm. The lower-energy state for all transitions is the ground state, which is $6s^2~^1S_0$ for Yb and $6s~^2S_{1/2}$ for Yb$^+$. The $4f$ orbital is fully filled with 14 electrons unless otherwise specified. $\Gamma$ is the full natural linewidth of the transition. For some transitions $\Gamma$ varies between isotopes.}
	\label{tableTransitions}
\begin{tabular}{llllll} 
 \hline
 \hline

Yb			&					& 			& Yb$^+$	&					& 		\\
$\lambda$	& excited state		& $\Gamma /2\pi$	& $\lambda$	& excited state		& $\Gamma /2\pi$	\\ 
 \hline
361			& $5d6s~^1D_2$		& 24 kHz	& 411		& $5d~^2D_{5/2}$	& 22 Hz	\\
399			& $6s6p~^1P_1$		& 29 MHz	& 436		& $5d~^2D_{3/2}$	& 3 Hz	\\
431			& $4f^{13}5d6s(J=2)$& 0.8 mHz	& 467		& $4f^{13}6s^2~^2F_{7/2}$	& 0.5 nHz	\\
556			& $6s6p~^3P_1$		& 184 kHz	& 			&	\\
578			& $6s6p~^3P_0$		& 7 mHz	& 			&	\\
 \hline
 \hline

\end{tabular}
\end{table}

In this paper, we report absolute frequencies for the 431-nm transition in $^{170,172,173,174,176}$Yb. This report completes the list of isotope shifts of the 431-nm transition for all stable isotopes at 10-kHz level. Based on these new measurements and some previously reported isotope shifts for other transitions, hyperfine structure, nuclear charge radii, and the King plot are analyzed. The King plot analysis also leads to a search for new bosons mediating additional force between a neutron and an electron. The results provide some insights on the nuclear structure for the odd-mass isotopes for Yb, motivating precise isotope-shift measurements for other transitions. 

\section{Experimental Measurements}
The experimental setup and sequences are previously reported elsewhere \cite{PhysRevA.107.L060801}. To summarize briefly, $\sim10^5$ atoms at $\sim30$ $\upmu$K are prepared in a magneto-optical trap (MOT) formed with the 556-nm transition. Atoms are then interrogated by 431-nm probe light at 10-mW power generated by second harmonic generation from the 862-nm light emitted by a titanium-sapphire laser. The frequency of the 862-nm light is stabilized to a ultralow-expansion cavity through a frequency comb and a 1064-nm laser, with all relevant radiofrequency signals referenced to the 10-MHz clock signal supplied from a physical realization of Coordinated Universal Time maintained by the National Metrology Institute of Japan. The 431-nm probe light is retroreflected, so that the atoms are immune to the Doppler shift. 

\begin{figure}[t]
    \includegraphics[width=1\columnwidth]{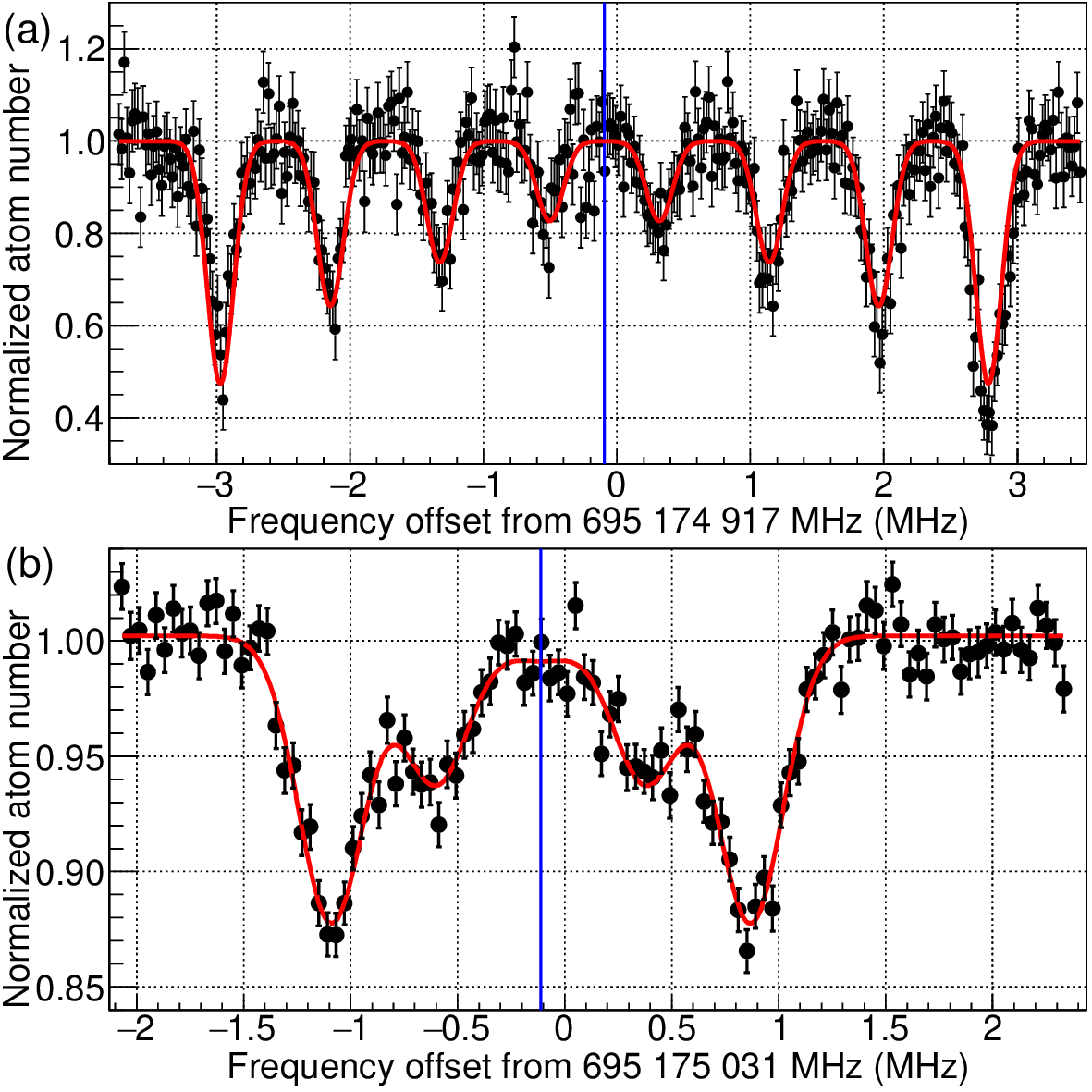}
    \caption{Spectra of the depletion of the MOT due to the 431-nm transition: (a) the $F=7/2$ hyperfine state in $^{173}$Yb with 0.0866(13)-mT bias magnetic field applied along the 431-nm probe beam. (b) $^{174}$Yb. Black points show the experimental data, and the red curve is the fit of the datapoints. The blue vertical line is the average frequency obtained from the fit. See the main text for the form of the fit function.} 
    \label{FigFixedFreqScan}
\end{figure}

The search for the 431-nm transition is first performed for each isotope by chirping the frequency of the 431-nm light while it is irradiated to the atoms in the MOT. The chirp rate varies from 200 kHz/s to 2 MHz/s depending on the expected strength of the transition. Once the initial signal of the transition is observed, further precise measurement of the resonant frequency is performed with the frequency of the 431-nm light fixed within a cycle. For the $F=7/2,~5/2$, and $3/2$ hyperfine states in $^{173}$Yb, the cycle of activating the 431-nm probe light for 3 ms while the MOT is turned off for a short period is repeated over 45 times. For other hyperfine states in $^{173}$Yb and even-mass isotopes, the 431-nm light is continuously irradiated onto atoms for 1 s without turning off the MOT. In both cases, the number of atoms after the irradiation of the 431-nm light normalized by the initial atom number is estimated from the amount of fluorescence from the MOT. Every time atoms are removed and then reloaded to the MOT, the frequency of the 431-nm light is shifted to obtain the spectrum of normalized atom number $R$.

Fig. \ref{FigFixedFreqScan} (b) and (a) show sample spectra for $^{174}$Yb and the $F=7/2$ state in $^{173}$Yb, respectively. For the $F=7/2$ state in $^{173}$Yb, distinct separation of Zeeman sublevels is observed due to a bias magnetic field of 0.0866(13) mT applied only when the MOT is deactivated, whereas the Zeeman sublevels in the $^{174}$Yb spectrum overlap because of the inability of applying arbitrary bias magnetic field with the MOT active. The spectra are fitted with the following equation:
\begin{equation}
R=R_0-\sum_{m=-F}^{F} A_{|m|}\exp \left[-\frac{\left(\nu-\left(\nu_0+m \Delta \nu\right)\right)^2}{2 \sigma^2}\right],
\end{equation}
where $R_0$ is a constant and $\sigma$ is the width for the Gaussian. The average frequency $\nu_0$ is regarded as the resonant frequency for the transition, and the amount of splitting $\Delta\nu$ is induced by the Zeeman shift. 

The absolute frequency of the transition is calculated by compensating systematic shifts from $\nu_0$. Major sources are the ac Stark shift due to the 431-nm probe beam and the Doppler shift. The former is estimated for $^{173}$Yb and the even-mass isotopes separately, due to the potential shift in the relative position between the MOT and the probe beam. The frequency shift per 1-mW probe beam is estimated to be 2.40(1.20) and 0.83(1.41) kHz for $^{173}$Yb and $^{174}$Yb as the representative of the even-mass isotopes, respectively. 

The first-order Doppler shift is canceled out by the retroreflection of the probe beam. The second-order Doppler shift for 30 $\upmu$K is negligible at the level of 1 kHz uncertainty. In the spectroscopy of a narrow-linewidth transition, a doublet of small recoil-free peaks is observable in the Doppler-broadened dip \cite{PhysRevLett.95.083003,PhysRevLett.101.183004}. In theory, the average frequency of this doublet and the bottom of the dip are at the same frequency, but in practice, these two frequencies do not agree. A 3.8 kHz disagreement is observed in our measurement \cite{PhysRevA.107.L060801}, and this is added to the uncertainty . For the even-mass isotopes, the frequency shift of 101.8(7.8) kHz induced by the ac Stark shift due to the MOT beam is also taken into account, as estimated in Ref. \cite{PhysRevA.107.L060801}. All other systematic shifts that are carefully measured in the $<1$ Hz level precision spectroscopies, such as second order Zeeman shift \cite{PhysRevA.98.022501}, Stark shift by ambient fields, collisional shift, blackbody radiation shift, and uncertainty of time base, are well below 1 kHz, and thus their contribution to the uncertainties at this level of accuracy is negligible. 

Obtained absolute frequencies are summarized in Table \ref{AbsFreqSummary}. As for $^{168}$Yb, which cannot be trapped with large enough number in the current setup, the absolute frequency is calculated from that for $^{174}$Yb and the isotope shift reported in Ref. \cite{PhysRevLett.130.153402}. Differences of the absolute frequencies between even-mass isotopes reported in this paper agree well within the uncertainty with the isotope shifts reported in Ref. \cite{PhysRevLett.130.153402}. 

For the $F=7/2$, $5/2$, and $3/2$ states in $^{173}$Yb, the Land\'e $g$ factor is measured by applying bias magnetic field along the direction of the incident probe beam (see Appendix \ref{AppendixA} for details). The bias magnetic field is calibrated using the Zeeman splitting of the 556 nm transition with a relative uncertainty of 1.5\%. The obtained $g$ factors are $g_F=0.662(46)$, $0.546(24)$, and $0.196(24)$ for $F=7/2$, $5/2$, and $3/2$, respectively. $g_J$ obtained by the weighted average of these $g_F$ is $1.583(56)$, which is consistent with the value in Ref. \cite{PhysRevA.107.L060801} within an uncertainty, but is off from theoretically expected $g_J=1.5$ and the value in Ref. \cite{PhysRevLett.130.153402}. 

\begin{table}[tb]
\centering
	\caption{Absolute frequencies and isotope shift $\nu_{431}^{A176}$ for the 431-nm transition in Yb isotopes: regarding $^{171}$Yb and $^{173}$Yb, the measured quantities are the transition frequencies for each $F$ state. The number listed in the line without $F$ represents the center of gravity of all $F$ states. }
	\label{AbsFreqSummary}
\begin{tabular}{lllrl} \toprule
$A$ 	& F		& Absolute frequency [kHz] 	& $\nu_{431}^{A176}$ [kHz]	& Reference \\ 
 \hline
 \hline
168		& 		& 695 170 466 295(17)		& -5 680 383(25)	& \cite{PhysRevLett.130.153402}		\\
170		& 		& 695 172 220 247(19)		& -3 926 431(26)	& this work	\\
171		& 		& 695 172 739 889(24)		& -3 406 789(30)	& 			\\
		& 3/2	& 695 171 054 858.1(8.2)	& 	& \cite{PhysRevA.107.L060801}	\\
		& 5/2	& 695 173 863 243(30)		& 	& \cite{PhysRevA.107.L060801}		\\
172		& 		& 695 173 850 275(18)		& -2 296 403(25)	& this work	\\
173		& 		& 695 174 348 875(19)		& -1 797 802(26)	& 			\\
		& 1/2	& 695 175 324 560(47)		& 					& this work	\\
		& 3/2	& 695 175 625 655(17)		& 					& this work	\\
		& 5/2	& 695 175 702 427(13)		& 					& this work	\\
		& 7/2	& 695 174 916 888(12)		& 					& this work	\\
		& 9/2	& 695 172 376 487(17)		& 					& this work	\\
174		& 		& 695 175 030 891(17)		& -1 115 787(25)	& this work	\\						
176		& 		& 695 176 146 678(17)		& 					& this work	\\
 \bottomrule
\end{tabular}
\end{table}

\section{Theoretical calculation of the electronic structure}\label{SecTheoryCalc}
The electronic structure of Yb is numerically calculated with configuration interaction implemented in {\textsc{amb}{\footnotesize i}\textsc{t}} \cite{CompPhysCommun.238.232}. For Yb, up to $n=19$ states for $s$ and $p$ orbitals and $n=14$ states for $d$ and $f$ orbitals are included in the calculation. For Yb$^+$, up to $n=16$ states are incorporated for all $s$, $p$, $d$, and $f$ orbitals. Two electron excitations from states relevant to this analysis and some other excited states and one hole excitation from the $4f$ orbital is allowed. For Yb, a hole excitation from the $5s$ and $5p$ orbitals is also allowed. $F_{\alpha}$, $K_{\alpha}$, and $D_{\alpha}$ are calculated by adjusting the initial conditions to observe the resulting energy shifts of each excited state. 

The results of our calculation are shown in Table \ref{tableTheoryCalc}, together with experimental measurements and results in other reports. Given the situation where sometimes $>50$\% disagreements are observed between other reports and experimental measurements (see $\nu_{467}$ in Table \ref{tableTheoryCalc}), most of our calculations agree well with experimental values and other reports, within $<10$\% relative differences. As for the 431-nm transition, $\nu_{431}$ and $F_{431}/F_{578}$ have experimental measurements. The relative offsets of the calculation from experiments for these two quantities are 17 and 13\%, respectively. The obtained mass shift has large differences to other calculations. However, previous reports on theoretical calculation also have large discrepancy between each other with very large relative uncertainties in some cases. Given this, it might be necessary to conservatively assume $\pm20$\% relative uncertainty, but our calculation is overall reasonably compatible with previous reports. 

\begin{table*}[tb]
\centering
	\caption{Result of the theoretical calculation of the electronic structure in Yb and Yb$^+$: the units of $\nu_{\alpha}$, $F_{\alpha}$, $K_{\alpha}$, $D_{\alpha}$, $K_{\beta,\alpha}$, and $D_{\beta,\alpha}$ are THz, GHz/fm$^2$, GHz u, $10^3$ THz, GHz u, and $10^3$ THz, respectively. $D_{\alpha}$ values are at $m_{\phi}=1$ eV. Experimental values for $\nu_{\alpha}$ are cited from Ref. \cite{NISTASD}, and the exact value differs by a few GHz according to isotope shifts. Experimental values for $F_{\beta}/F_{\alpha}$ and $K_{\beta\alpha}$ are obtained by the linear fit of King plots for the even-mass isotopes (see Section \ref{secking}). In Refs. \cite{PhysRevA.49.3351,AtomDataNuclData.99.69,NuclChargeRadii}, the convention of the mass shift is different and a negative sign is added to compensate it.}
	\label{tableTheoryCalc}
\begin{tabular}{lrrrrrrrr} \toprule
\multicolumn{2}{r}{this work}	& {\textsc{amb}{\footnotesize i}\textsc{t}} \cite{PhysRevLett.128.163201}& {\textsc{grasp}2018} \cite{PhysRevLett.128.163201}& {\textsc{grasp}2018} \cite{PhysRevA.104.022806}			&Ref. \cite{AtomDataNuclData.99.69}&Ref. \cite{NuclChargeRadii}& Other reports		& Experiment  \\
 \hline
 \hline
$\nu_{361}$	& 875.19	& 819.47	& 			& 			&&&	& 829.76 \\
$\nu_{399}$	& 744.82	& 			& 			& 			&&&	& 751.53 \\
$\nu_{411}$	& 691.88	& 707.00	& 808.11	& 			&&&	& 729.48 \\
$\nu_{431}$	& 810.67	& 			& 			& 			&&&	& 695.17 \\
$\nu_{436}$	& 649.86	& 679.86	& 770.13	& 			&&&	& 688.35 \\
$\nu_{467}$	& 727.06	& 1051.44	& 580.12	& 			&&&	& 642.12 \\
$\nu_{556}$	& 524.06	& 			& 			& 543.18	&&&	& 539.39 \\
$\nu_{578}$	& 500.53	& 522.78	& 458.36	& 522.68	&&&	& 518.29 \\
 \hline
$F_{361}$	& -11.988	& -13.528	& 			& 			&&& -14.437 \cite{PhysRevLett.128.073001}	&  \\
$F_{399}$	& -7.020	& 			& 			& 			&&&	&  \\
$F_{411}$	& -15.534	& -14.715	& -15.852	& 			&&& -17.604 \cite{PhysRevLett.128.073001}	&  \\
$F_{431}$	& 17.771	& 			& 			& 			&&&	&  \\
$F_{436}$	& -15.770	& -14.968	& -16.094	& 			&&& -18.003 \cite{PhysRevLett.128.073001}	&  \\
$F_{467}$	& 35.260	& 36.218	& 41.892	& 			&&&	&  \\
$F_{556}$	& -9.893	& 			& 			& -10.951(21)	&-11.5 &-9.3(2.1) & -12.3(2) \cite{PhysRevA.49.3351}	&  \\
$F_{578}$	& -9.692	& -9.719	& -9.1508	& -10.848(21)	&&&	&  \\
 \hline
$K_{399}$	& 412		& 			& 			& 			&&&	&  \\
$K_{411}$	& -308		& -752		& -1678.2	& 			&&&	&	  \\
$K_{431}$	& 10172		& 			& 			& 			&&&	&  \\
$K_{436}$	& -158		& -661		& -1638.5	& 			&&&	&  \\
$K_{467}$	& 10167		& 12001		& 3127.6	& 			&&&	&  \\
$K_{556}$	& -532		& 			& 			& -280(72)	&-4581(1531) &-580(3820) & -1500(500) \cite{PhysRevA.49.3351}	&  \\
$K_{578}$	& -527		& 			& 			& -288(75)	&&& -655 \cite{PhysRevResearch.2.043444}\hspace{1.72mm}	&  \\
 \hline	
$D_{411}$	& 43.37		& 43.158	& 44.145	& 			&&& 41.235 \cite{PhysRevLett.128.073001}	&  \\
$D_{431}$	& -194.34	& 			& 			& 			&&&	& 	\\
$D_{436}$	& 49.74		& 48.634	& 48.419	& 			&&& 48.795 \cite{PhysRevLett.128.073001}	&  \\
$D_{467}$	& -295.06	& -352.38	& -730.4	& 			&&&	&  \\
$D_{578}$	& -39.49	& -42.855	& -55.729	& 			&&&	&  \\
\hline
$F_{399}/F_{578}$	& 0.7243	& 	& 			& 			&&&	& 0.4453(12) \\
$F_{431}/F_{578}$	& -1.8336	& 	& 			& 			&&&	& -1.62301(8) \\
$F_{556}/F_{578}$	& 1.0207	& 	& 			& 1.0095(28)&&&	& 1.00594(26) \\
$K_{399,578}$		& 793.7		& 	& 			& 			&&&	& 1332(17) \\
$K_{431,578}$		& 9205.7	& 	& 			& 			&&&	& 6430.9(1.4) \\
$K_{556,578}$		& 5.929		& 	& 			& 10.7(104.5)	&&&	& 33.2(4.0) \\
$D_{431,578}$		& -266.7	& 	& 			& 			&&&	&  \\

 \bottomrule
 
\end{tabular}
\end{table*}

\section{Data Analysis}
\subsection{Hyperfine Structure}
From the absolute frequencies listed in Table \ref{AbsFreqSummary}, hyperfine constants are calculated. The hyperfine A constant $A_h^{171}$ for $^{171}$Yb is reported in Ref. \cite{PhysRevA.107.L060801}. For $^{173}$Yb, first-order hyperfine interaction is characterized by the A, B, and C hyperfine constants $A_h^{173}$, $B_h^{173}$, and $C_h^{173}$ corresponding to the magnetic dipole, electric quadrupole, and magnetic octupole components (see Appendix \ref{AppendixB}). The best fits of the absolute frequencies with and without the octupole term are summarized in Table \ref{table173HF}. With the octupole term, the fit quality characterized by $\chi^2/{\rm ndf}$ (ndf: number of degrees of freedom) improves but is still somewhat poor with $\chi^2/{\rm ndf}=6.711$. The obtained octupole moment $C^{173}_h=7.6(4.7)$ kHz, which includes the inflation of the uncertainty with $\sqrt{\chi^2/ {\rm ndf}}$, has only a significance $1.62\sigma$, and the shift of the dipole and quadrupole constants induced by the introduction of the octupole term is within the standard deviation after inflating it by $\sqrt{\chi^2/ {\rm ndf}}$. 

The effect of the second-order hyperfine interaction $W_F^{(2)}$ can also be significant in Yb (see Appendix \ref{AppendixB} for the formalism). To estimate its effect, another fit is performed with $A_h^{173}$, $B_h^{173}$, $\eta$, and $\zeta$. The result is summarized in the right column of Table \ref{table173HF}. The A and B constants are consistent with the fit only with the A and B constants within a standard deviation inflated by $\chi^2/{\rm ndf}$. The amount of the second-order hyperfine interaction contributing to $A_h^{173}$, $B_h^{173}$, and $C_h^{173}$ is $5.1\pm 1.0$, $-22\pm15$, and $-7.6\pm 1.8$ kHz, respectively. At the present accuracy of the spectroscopy, these numbers are comparable to the uncertainty with the inflation by $\sqrt{\chi^2/ {\rm ndf}}$. Further improvement in the accuracy of the spectroscopy and inputs from theoretical calculations is desired to precisely determine the amount of second-order hyperfine interaction and whether the nucleus has a finite octupole moment. Note that same $\chi^2$ for the second and third column in Table \ref{table173HF} is not a coincidence. This is because the dependence of the second-order hyperfine interaction when $\zeta =-3\eta$ is the same as a linear combination of $\nu_{\rm ave}$, $A^{173}_h$, $B^{173}_h$, and $C^{173}_h$. The second-order hyperfine interaction shifts the center-of-mass frequency by $\-3.7(5.0)$ kHz, which is smaller than the uncertainty listed in Table \ref{AbsFreqSummary}. This is not large enough to explain the residual of the datapoint for $^{173}$Yb from the even-mass isotopes' linearity in the King plot (See Sec. \ref{secking}). It should be noted that in Ref. \cite{PhysRevA.104.022806}, analysis for $6s6p~^3P$ states based on a theoretical calculation of the wave function gives $\sim-500$-kHz shift for the center-of-mass frequency. It is possible that with such calculations, larger shift is predicted for the 431-nm transition as well. Further investigation on the effect of the second-order hyperfine structure on the shift in the center-of-mass frequency is necessary. In the following analysis, to be conservative, $A_h^{173}$ and $B_h^{173}$ obtained by the fit without $C^{173}_h$, $\eta$ and $\zeta$ are used. 

\begin{table}[t]
\centering
	\caption{Hyperfine constants for odd-mass Yb isotopes in the unit of kHz: the top row specifies parameters included in the fit in each column. The uncertainty for $^{173}$Yb does not include the inflation by $\sqrt{\chi^2/ {\rm ndf}}$. $A^{171}_h$ is cited from Ref. \cite{PhysRevA.107.L060801}.}
	\label{table173HF}
\begin{tabular}{llll} \toprule
	 					& $A_h,~B_h$		& $A_h,~B_h,~C_h$	& $A_h,~B_h,~\eta,~\zeta$  \\ 
 \hline
 \hline
$^{171}$Yb $A^{171}_h$~	& 1~123~354(13)~	& 					& 					 \\
$^{173}$Yb $A^{173}_h$~	& -309~449.1(2.0)~	& -309~446.1(2.1)	& -309~441.0(2.6)			\\
		   $B^{173}_h$~	& -1~700~631(22)~	& -1~700~652(23)	& -1~700~674(27)	 \\
		   $C^{173}_h$~	& 					& 7.6(1.8)	 		& 						\\
		$\eta^{173}_h$~	&					& 			 		& 1~129(339)			\\
		$\zeta^{173}_h$~&					& 	 				& -3~774(895)			\\
$\chi^2/{\rm ndf}$		& 24.1/2			& 6.711/1			& 6.711/0				\\	
 \bottomrule
\end{tabular}
\end{table}

The ratio of $A_h$ between different isotopes equals the ratio of the nuclear $g$ factors $g^A$, as long as a pointlike nuclear dipole moment is assumed. The deviation $\Delta_{A,A'}$ defined by the following equation is referred to as hyperfine anomaly \cite{RevModPhys.49.31}: 
\begin{equation}
\Delta_{171,173}=\frac{A_h^{171}/A_h^{173}}{g^{173}/g^{171}}-1=\frac{A_h^{171}/A_h^{173}}{(\mu_n^{173}/I^{173})/(\mu_n^{171}/I^{171})}-1,
\end{equation}
where $\mu_n^A$ and $I^A$ are the nuclear magnetic moment and nuclear spin, and $(\mu_n^{171}/I^{171})/(\mu_n^{173}/I^{173})=-3.6305(1)$ \cite{ZPhys.249.205,PhysRevA.49.3351,AtomDataNuclDataTable.90.75}. The hyperfine anomalies for the 431-nm and other transitions are summarized in Table \ref{tableHFanomaly}. $\Delta_{171,173}$ for the $4f^{13}5d6s^2(J=2)$ state corresponding to the 431-nm transition is finite but particularly small compared to other states in Yb, as well as anomalies in most alkali-metal atoms \cite{RevModPhys.49.31}. Also, it is worth noting that the singlet states tend to have large anomaly. 

\begin{table}[t]
\centering
	\caption{Hyperfine anomalies $\Delta_{171,173}$ for different states in Yb: references provide values for the A and B constants.}
	\label{tableHFanomaly}
\begin{tabular}{lll} \toprule
State	 	&	$\Delta_{171,173}/10^{-3}$	& Ref. \\ 
 \hline
 \hline
$6s6p~^1P_1$			& 13.90(56)			& \cite{PhysRevA.72.032506} \\
$4f^{13}5d6s^2(J=2)$	& -0.084(55)		& this work		\\
$6s6p~^3P_2$			& -8.96(53)			& \cite{JPSJ.72.2219}	\\
$6s6p~^3P_1$			& -3.856(70)		& \cite{PhysRevA.100.042505} \\
$6s7s~^3S_1$			& -3.50(34)			& \cite{JPSJ.72.2219}	\\
$6s5d~^1D_2$			& -13.0(6.7)		& \cite{JOSAB.3.332}	\\	
 \bottomrule
\end{tabular}
\end{table}

\subsection{Nuclear charge radius}
\begin{table}[!t]
\centering
	\caption{Summary of $\lambda ^{AA'}$ and $\delta\langle r^2\rangle ^{AA'}$ with $A'=176$: three digit numbers specify the transition used to obtain $\lambda ^{AA'}$ or $\delta\langle r^2\rangle ^{AA'}$. The columns without references are based on our data, except that the 578 isotope-shift data are cited from Ref. \cite{PhysRevX.12.021033}. }
	\label{tableRMSradii}
\begin{tabular}{l|ll|llll} \toprule
$A$	&	\multicolumn{2}{c|}{$\lambda ^{AA'}$ (fm$^2$)}	& \multicolumn{4}{c}{$\delta\langle r^2\rangle ^{AA'}$ (fm$^2$)~~~~} \\ 
	&	431 	& x-ray \cite{PhysRevA.20.239}	& 431  	& 578 	& 556  \cite{PhysRevA.104.022806}	& Ref. \cite{NuclChargeRadii} \\ 
 \hline
 \hline	
168	& -0.475	& 				& -0.505	& -0.517	& -0.4548(21)		& -0.550(128) 	\\
170	& -0.336	& -0.407(26)	& -0.357	& -0.364	& -0.3200(16)	 	& -0.392(38) 	\\
171	& -0.287	& -0.330(23)	& -0.305	& -0.312	& -0.2744(13)	 	& -0.329(77)	\\
172	& -0.205	& -0.244(18)	& -0.218	& -0.220	& -0.1934(10)	 	& -0.233(24)	 \\
173	& -0.158	& -0.194(35)	& -0.168	& -0.170	& -0.1493(8)	 	& -0.180(24) 	\\
174	& -0.100	& -0.103(12)	& -0.107	& -0.108	& -0.0944(5)	 	& -0.117(30) 	\\
 \bottomrule
\end{tabular}
\end{table}

With theoretically calculated $F_{\alpha}$ and $K_{\alpha}$, differences in nuclear charge radii can be estimated using only the first two terms in Eq. (\ref{eqnIS}). 
In this case where all higher order effects of nuclear charge are renormalized in $\delta\langle r^2\rangle^{AA'}$, the parameter is referred to as the nuclear charge parameter, and denoted as $\lambda ^{AA'}$. The obtained numbers based on the 431-nm transition data are shown in Table \ref{tableRMSradii}.  The relative uncertainty of $\delta\langle r^2\rangle^{AA'}$ caused by the uncertainty of the spectroscopy is $\sim10^{-5}$, and the uncertainty for the obtained $\delta\langle r^2\rangle^{AA'}$ mainly originates from that of the theoretical calculation of $F_{\alpha}$ and $K_{\alpha}$, the amount of which is difficult to accurately estimate and therefore not shown in the table. Note that a conservative estimate discussed in Section \ref{SecTheoryCalc} is $\pm 20$\%. 

To extract $\delta\langle r^2\rangle^{AA'}$ from $\lambda ^{AA'}$, a relation $\delta\langle r^2\rangle^{AA'}=\lambda ^{AA'}/0.941$ \cite{PhysRevA.104.022806,NuclChargeRadii} is utilized. The obtained $\delta\langle r^2\rangle ^{AA'}$ are shown in Table \ref{tableRMSradii} within the uncertainty of the combined analysis in Ref. \cite{NuclChargeRadii}, but they tend to be smaller. One way to explain this is the discrepancy in $F_{431}/F_{578}$ between the theory and experiment. Empirically correcting $F_{431}$ to match the theoretical value of $F_{431}/F_{578}$ to the experimental value increases $\delta\langle r^2\rangle ^{AA'}$ by 13\%. This reduces the discrepancy to Ref. \cite{NuclChargeRadii}, assuming that $F_{578}$ is reliable. Because $F_{578}$ and $K_{578}$ have a reasonable agreement between other reports, this assumption is reasonable to some extent, but the resulting $\delta\langle r^2\rangle ^{AA'}$ calculated by $F_{578}$ and $K_{578}$ obtained by our theory values and isotope shift in Ref. \cite{PhysRevX.12.021033} shown in Table \ref{tableRMSradii} is also smaller than Ref. \cite{NuclChargeRadii}. 

In fact, such smaller $\delta\langle r^2\rangle ^{AA'}$ derived from optical spectroscopies compared to combined analysis like Refs. \cite{NuclChargeRadii,AtomDataNuclData.99.69} is observed in other reports as well \cite{PhysRevA.104.022806,PhysRevA.49.3351,PhysRevA.20.239}. Values obtained in Ref. \cite{PhysRevA.104.022806}, which is relatively new and has a careful assessment of their theoretical calculation within their own system, is shown in Table \ref{tableRMSradii} as an example. The discrepancy seems to be because the x-ray spectroscopy provides relatively large $\lambda ^{AA'}$, as shown in Table \ref{tableRMSradii} \cite{PhysRevA.20.239}. Spectroscopy of muonic atoms, which is another major contributor to the combined analysis, gives a close number to the calculation by the 431-nm transition data \cite{PhysRevA.20.239}. Some scattering experiments also favor $|\delta\langle r^2\rangle ^{174,176}|$ smaller than 0.1 \cite{SasanumaThesis,CreswellThesis,PhysRevC.13.1083,PhysRevC.29.1228}. For further interpretation of the data, a combined analysis with recent data is desired.

\begin{figure}[t]
    \includegraphics[width=1\columnwidth]{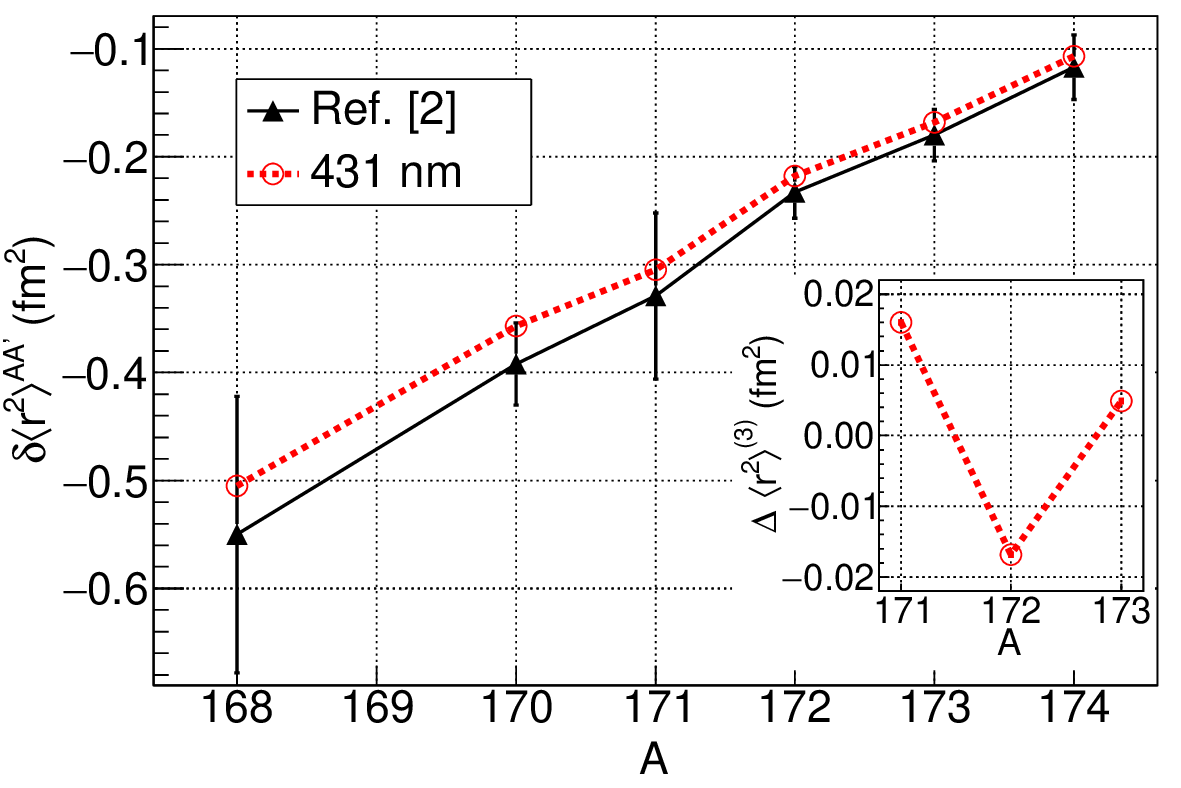}
    \caption{$\delta\langle r^2\rangle^{AA'}$ calculated from isotope shifts with $A'=176$: red circles show the values obtained from the absolute frequencies of the 431 nm transition, and black triangles show the values reported in Ref. \cite{NuclChargeRadii}. The inset shows $\Delta \langle r^2\rangle^{(3)}$. } 
    \label{FigRrms}
\end{figure}

From these data, three-point odd-even staggering $\Delta \langle r^2\rangle^{(3)}$ is calculated as
\begin{equation}
\Delta \langle r^2\rangle^{(3)}=\frac{1}{2} (\langle r^2\rangle^{A+1} - 2 \langle r^2\rangle^{A} + \langle r^2\rangle^{A-1} ).
\end{equation}
This is plotted in the inset of Fig. \ref{FigRrms}. Compared to the odd-even staggering observed in other atoms \cite{PhysRevC.105.014325,PhysRevLett.126.032502,PhysLettB.797.134805,NatPhys.16.620}, this effect is relatively large for stable isotopes. This could be attributed to the fact that both the number of protons and neutrons are far from the magic numbers, 50 and 82.

\subsection{King Plot}\label{secking}
\begin{figure}[t]
    \includegraphics[width=1\columnwidth]{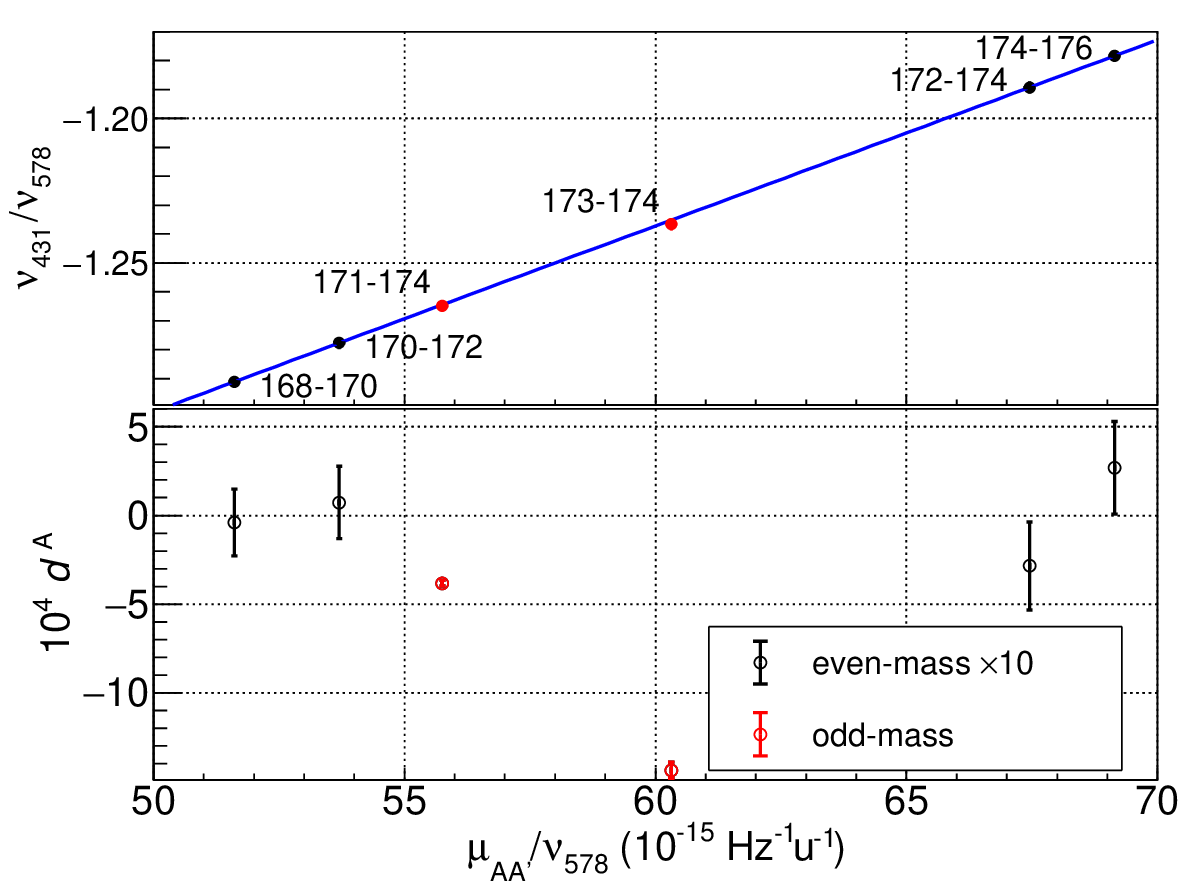}
    \caption{Normalized King plot for the 431-nm transition: the plot is normalized with the 578-nm transition. Text near each data point shows the pair of isotopes that generated the data point. The blue line shows the linear fit of the black data points. See Table \ref{tableTheoryCalc} for the fit parameters. The bottom half displays the residual $d^A$ of each data point from the fitted line. The residuals for the even-mass isotopes (black points) are magnified by a factor of 10.} 
    \label{FigKingPlot}
\end{figure}

The normalized King plot between the 431-nm transition and the 578-nm transition following Eq. \ref{eqnNormKing} is shown in Fig. \ref{FigKingPlot}. In addition to the data points for even-mass isotopes, data points for odd-mass isotopes are plotted. The representative transition frequency for an odd-mass isotope is calculated as the center of gravity frequency of the hyperfine levels, and the isotope shift is calculated as the shift from the transition frequency for $^{174}$Yb, so that the data points lie between those of even-mass isotopes, and because $^{174}$Yb has smaller uncertainties compared to other even-mass isotopes. To assess the level of nonlinearity of the King plot, linear fit for the even-mass isotopes is performed by Eq. \ref{eqnNormKing} with only the first two terms on the right-hand side. The plot exhibits slight nonlinearity beyond the standard deviation for some data points, as characterized by $\chi^2/{\rm ndf}=1.264$. However, as the probability for this level of deviation from $\chi^2/{\rm ndf}=1$ is 28.24\%, the data points are linear within 95\% confidence level. 

\begin{figure}[t]
    \includegraphics[width=0.8\columnwidth]{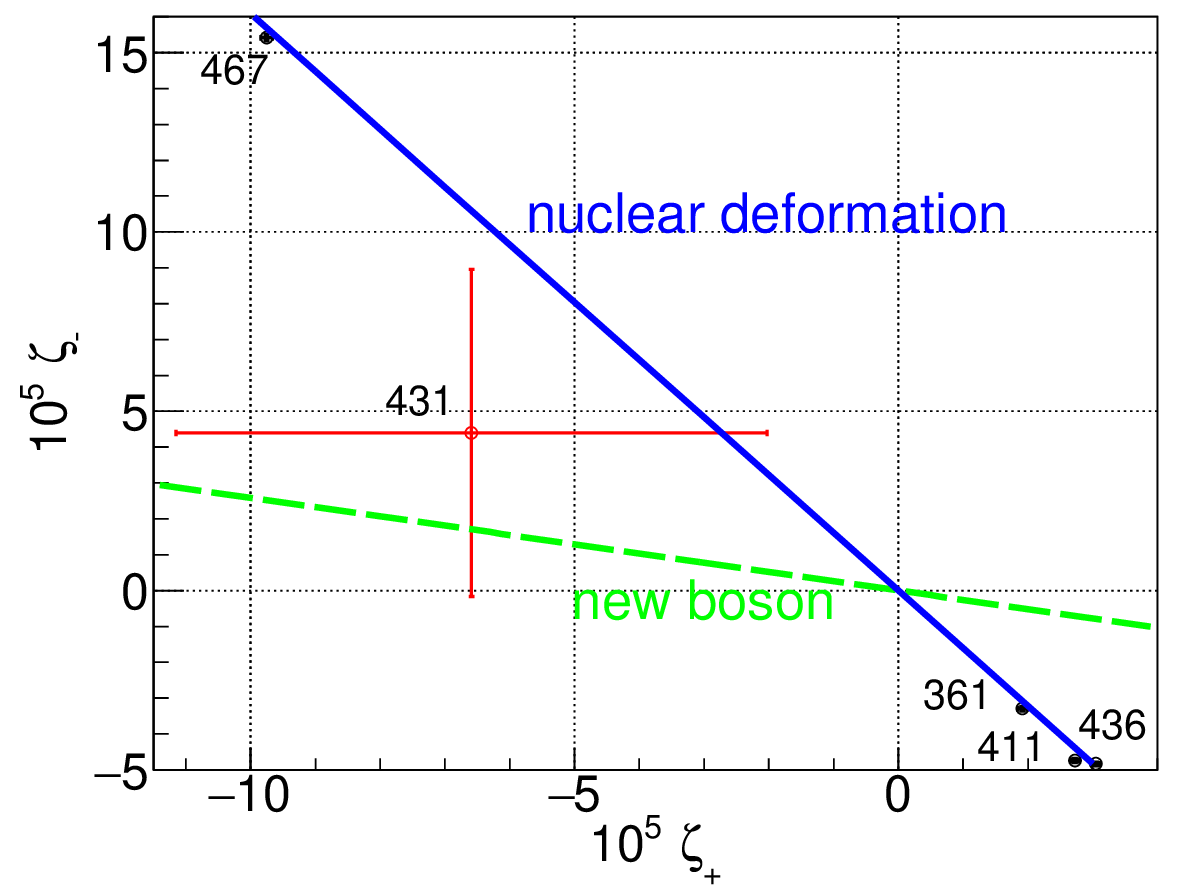}
    \caption{Observed nonlinearity in the normalized King plot displayed on the $(\zeta_+,\zeta_-)$ basis. The blue line is the linear fit for the black points with a line on the origin. Best fit is $\zeta_-=-1.609(14)\zeta_+$, with $\chi^2/{\rm ndf}=5.826$. The green dashed line is the nonlinearity pattern generated by the new bosons. } 
    \label{FigXiPM}
\end{figure}

Although the amount of the nonlinearity is consistent with zero, the residual $d^A$ for isotope $A$ can be decomposed onto the two-dimensional space made by $\zeta_{\pm}=d^{168}-d^{170}\pm(d^{172}-d^{174})$, as introduced in Ref. \cite{PhysRevLett.125.123002}. The nonlinearity of the 431-nm transition shown in Fig. \ref{FigXiPM} is consistent with the one-parameter fit for the 361-, 411-, 436- and 467-nm transitions within one standard deviation. This finding supports the argument made in Ref.~\cite{PhysRevLett.128.163201} that the major part of the nonlinearities in the King plots for Yb originates from several nuclear effects. This consistency observed between the previous results, as well as the almost linear behavior on the King plot, ensures that the measurements are not significantly affected by any systematic shifts beyond the listed uncertainties. 

\begin{table}[t]
\centering
	\caption{Residual $\delta\nu^{A}$ from the linear fit of the King plot for odd-mass isotopes: the residual on the plot is multiplied by $\nu^{AA'}_{578}$. $\lambda$ is the wavelength of the transition.}
	\label{tableOffset}
\begin{tabular}{lrr} \toprule
$\lambda$ (nm)	&	$\delta\nu^{171}$ (kHz)	& $\delta\nu^{173}$ (kHz) \\ 
 \hline
 \hline
399			& 380(218)		& 3302(266)	\\
431			& -692(29)~~	& -792(25)~	\\
556			& 1233(79)~~	&  455(76)~	\\
 \bottomrule
\end{tabular}
\end{table}

$d^A$ for the odd-mass isotopes is significantly larger than the uncertainties on it, as shown in the bottom half of Fig. \ref{FigKingPlot}. To investigate  the origin of this offset, the normalized King plots are made for the 399- and 556-nm transitions as well, and $\delta\nu^A=d^A \nu_{578}^{AA'}$ is extracted, as shown in Table \ref{tableOffset}. On one hand, it is at most $\sim1$ MHz and thus not significant in older experiments with large uncertainties \cite{JOSAB.3.332,PhysRevA.63.023402,JOSAB.11.2163}. On the other hand, with an accuracy of $\sim10$ kHz, this is significant. $\delta\nu^{171}$, $\delta\nu^{173}$, and $\delta\nu^{173}/\delta\nu^{171}$ are plotted against $\Delta_{171,173}/10^{-3}$, $A_h^{171}$, $B_h^{173}$, $B_h^{173}/A_h^{173}$, and $F_{\alpha}$. Quantities with largest correlation are $B^{173}$ for $\delta\nu_{173}$ and $F_{\alpha}$ for $\delta\nu_{171}$, respectively. $\delta\nu^{173}/\delta\nu^{173}$ does not appear to have any good correlation against other quantities. However, it is not evident enough to state these quantities are related, because the $\chi^2/{\rm ndf}$ for the linear fit are 7.329 and 5.634, respectively, and both fitted lines do not cross the origin on the two-dimensional plot. One reason why the analysis is inconclusive is because $\delta\nu^{171}$ for the 399-nm transition has a large relative uncertainty. More precise spectroscopy, not only limited to the 399-, 431-, and 556-nm transitions but also for other narrow-linewidth transitions, is desired to figure out the source of the large $d^A$ for odd-mass isotopes. 

\subsection{Search for new bosons}
\begin{figure}[t]
    \includegraphics[width=1\columnwidth]{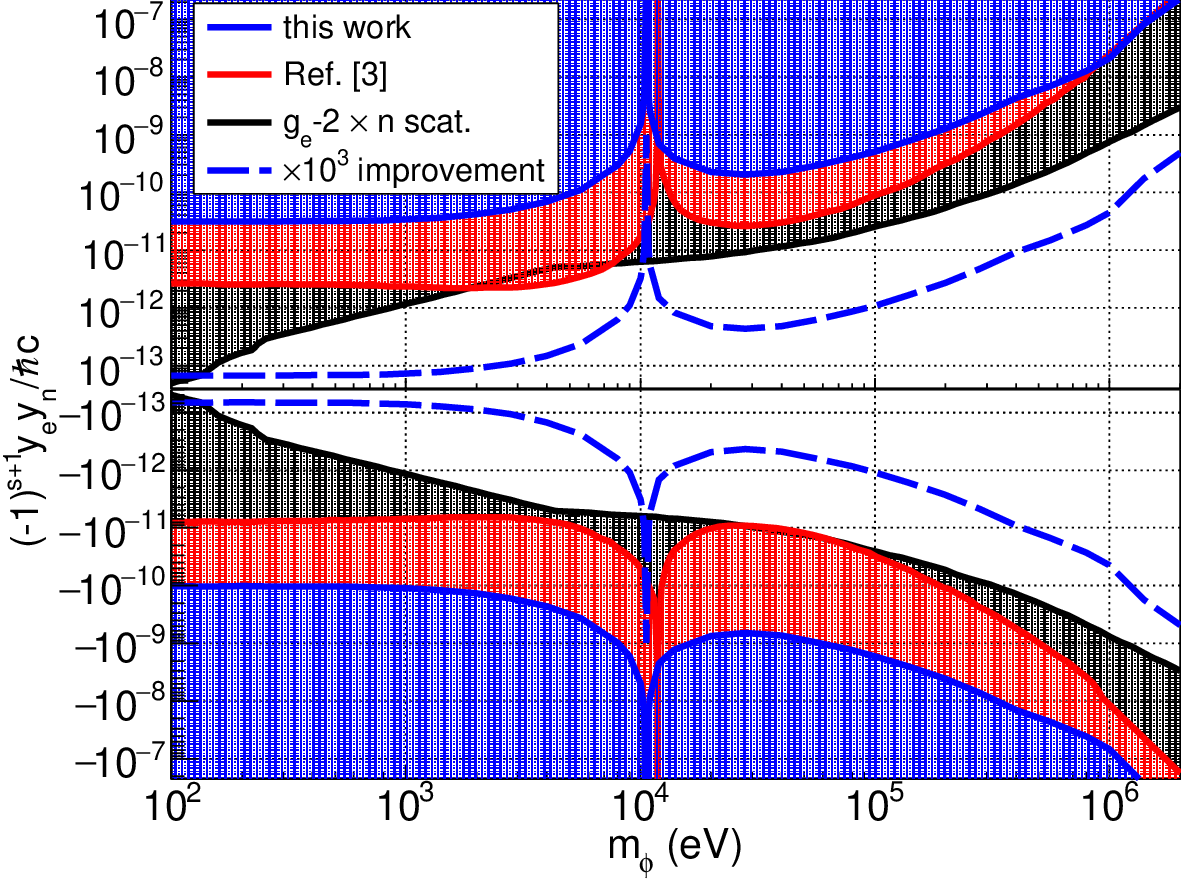}
    \caption{Constraint on the existence of new bosons mediating the force between an electron and a neutron: the black region is excluded by electron $g-2$ measurements and neutron-scattering experiments \cite{PhysRevLett.120.091801}. The red region is the most stringent exclusion reported so far in Ref. \cite{PhysRevLett.128.163201}, obtained from the analysis with the 436-, 467-, and 578-nm transitions. The blue region is the excluded region by the 431- and 578-nm transitions obtained in this work. The blue dashed line is the expected sensitivity when the accuracy in isotope shifts for 431 nm is improved by a factor of 1000.} 
    \label{FigNewBoson}
\end{figure}

Together with theoretically calculated $D_{\alpha}$ and the linearity in the King plot for the 431-nm transition, a constraint on the existence of new bosons mediating the force between an electron and a neutron is estimated. It should be noted that with the relative uncertainty of $\sim1\times10^{-5}$ for the isotope shifts in the 431-nm transition, simple analysis with a two-dimensional King plot is good enough, as other sources of uncertainties such as the ones on $\mu^{AA'}$ are orders of magnitude smaller. To be most conservative, the effect of the nuclear deformation along the blue line in Fig. \ref{FigXiPM} is first subtracted from the data point for the 431-nm transition, so that all nonlinearity would arise from the new bosons. As shown in Fig. \ref{FigXiPM}, the effect of the new boson on the $\zeta_+-\zeta_-$ plane has a slope of -0.2585. On this line, $-6.685\times 10^{5}<v_{ne} D_{431,578}<2.182 \times 10^{6}$ is the allowed region for the new bosons with 95\% confidence level, which includes $v_{ne}=0$. Thus, such bosons are excluded in the region beyond this, as plotted in Fig. \ref{FigNewBoson}. The constraint is more than an order of magnitude worse compared to the most stringent constraint reported in Ref. \cite{PhysRevLett.128.163201}. 

An improved accuracy of the isotope-shift measurement for the 431-nm transition increases the sensitivity to the region that has never been investigated before. Fig. \ref{FigNewBoson} shows the sensitivity for the improved accuracy by three orders of magnitude (down to $\sim18$ Hz), with an assumption that the uncertainty in the isotope shifts for the 431-nm transition is the primary source of the uncertainty. At this level, analysis with three transitions described in Ref. \cite{PhysRevResearch.2.043444} is necessary. 

\section{Conclusion}
The full list of absolute frequencies for the 431-nm transition in $^{170,172,173,174,176}$Yb is reported, and various analyses based on the data are performed. The King plot analysis did not show as much nonlinearity as reported for other transitions. On one hand, this can ensure the absence of any significant systematic shifts in the measurement, but on the other hand, further improvement in accuracy is desired to investigate the effects inducing the nonlinearity, including the new bosons mediating the force between an electron and a neutron. The hyperfine constants are precisely determined. However, the observed octupole moment is not statistically significant, and this also motivates more precise spectroscopy. The difference in root-mean-square nuclear charge radii between isotopes determined by our analysis supports smaller values compared to some of the previous reports. 

\begin{acknowledgments}
This work was supported by Japan Society for the Promotion of Science KAKENHI Grants No. JP21K20359, JP22H01161, JP22K04942, and Japan Science and Technology Agency FOREST Grant No. JPMJFR212S and JST-MIRAI Grant No. JPMJM118A1. We are grateful to D. Akamatsu, K. Hosaka, H. Inaba, and S. Okubo for the development of the frequency comb and the stable laser at 1064 nm. A. K. acknowledges the partial support of a William M. and Jane D. Fairbank Postdoctoral Fellowship of Stanford University for the early stage of this project. 
\end{acknowledgments}

\appendix
\begin{figure*}
    \subfloat{\includegraphics[width=0.32\textwidth]{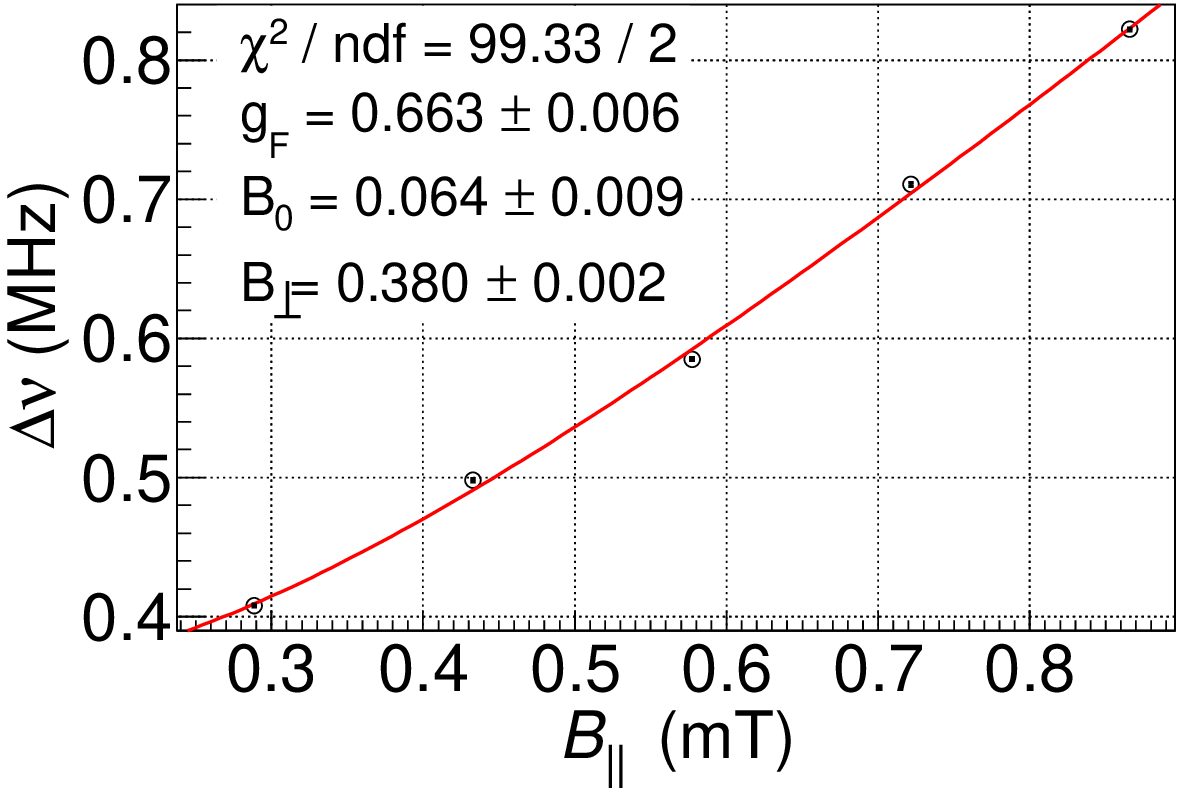}} 
    \subfloat{\includegraphics[width=0.32\textwidth]{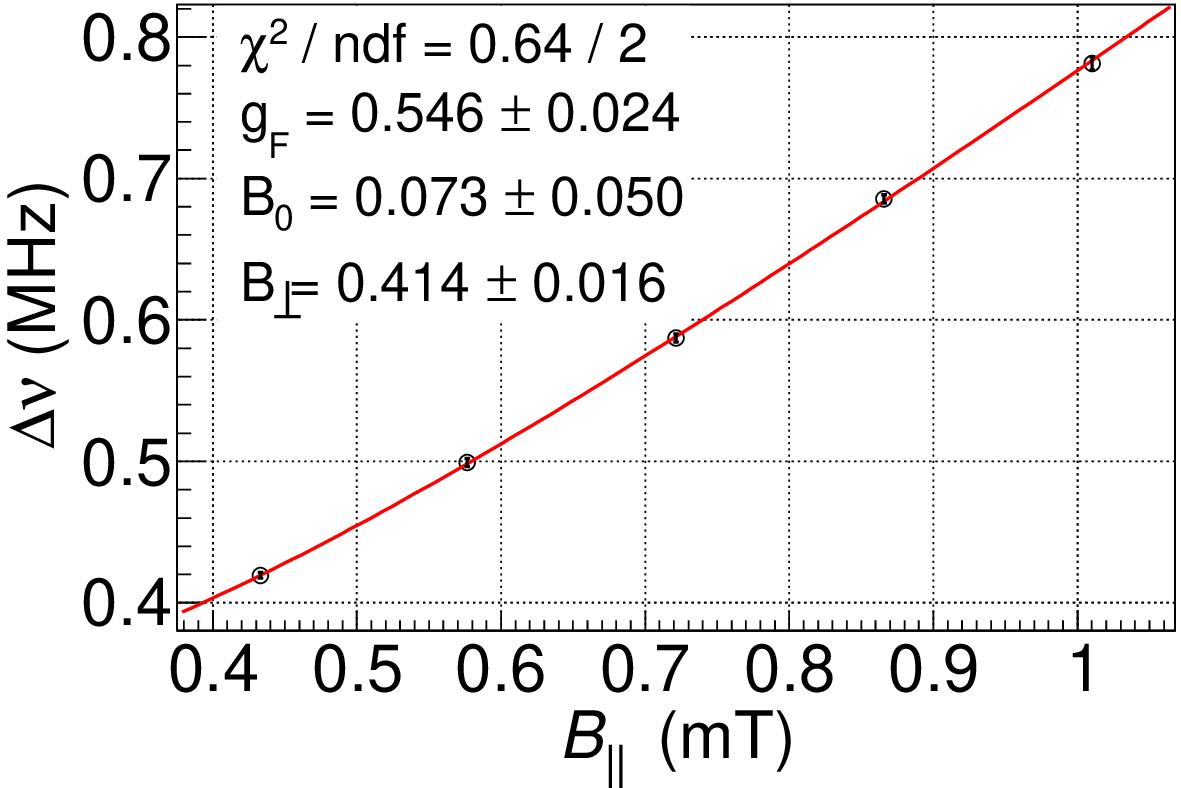}} 
    \subfloat{\includegraphics[width=0.32\textwidth]{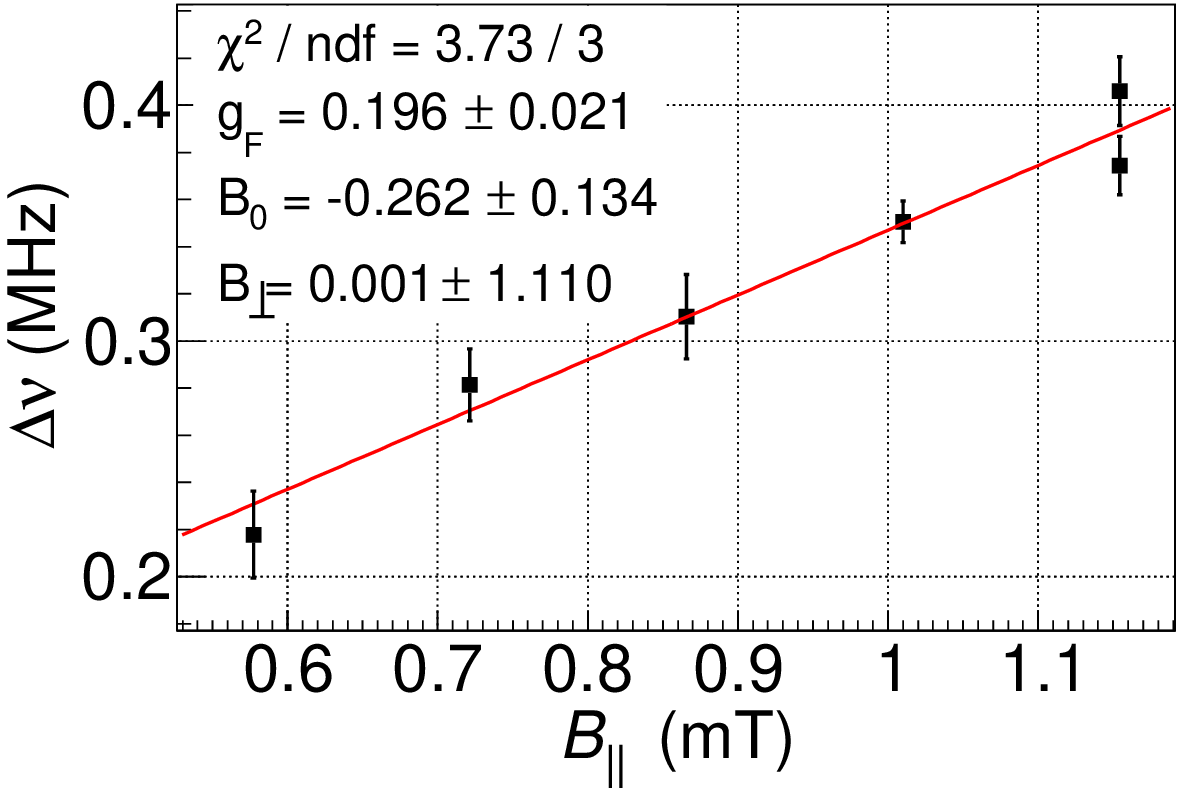}} 
   \caption{The amount of Zeeman splitting $\Delta \nu$ for applied bias magnetic field $B_{\parallel}$ for the (a) $F=7/2$, (b) $F=5/2$, and (c) $F=3/2$ hyperfine states in $^{173}$Yb. The uncertainties on the final values for $g_F$ in the main text are inflated by $\sqrt{\chi^2/{\rm ndf}}$. if $\sqrt{\chi^2/{\rm ndf}}>1$. }
    \label{FigZeeman}
\end{figure*}

\section{Details of the $g$ factor determination}\label{AppendixA}
To determine the $g$ factors for the $F=7/2$, $5/2$, and $3/2$ states in $^{173}$Yb, the bias magnetic field $B_{\parallel}$ along the direction of the propagation of the 431-nm probe beam is varied to obtain the amount of Zeeman splitting. These are shown in Fig. \ref{FigZeeman}. The plot is fitted with the fit function $\Delta \nu= g_F \mu_B \sqrt{(B_\parallel-B_0)^2+B_\perp^2}$ to accommodate the transverse magnetic field $B_\perp$, where $g_F$, $B_{0}$, and $B_\perp$ are fitting parameters. $g_F$ is converted to $g_J$ with the relation 
\begin{equation}
g_J\simeq \frac{2F(F+1)}{F(F+1)-I(I+1)+J(J+1)}g_F.
\end{equation}
With the uncertainty obtained by the fit, the contribution of the nuclear spin to $g_F$ is negligible.

\section{Formula for hyperfine structure}\label{AppendixB}
To the extent of the first-order interaction, the absolute frequency $\nu_F$ for the hyperfine level with total angular momentum $F$ in $^{173}$Yb satisfies the following equation. 
\begin{widetext}
\begin{eqnarray}\label{EqHFABC}
\nu_F&=&\nu_{\rm ave}+A_h^{173}\frac{K}{2} + B_h^{173}\frac{3K(K+1)-4I(I+1)J(J+1)}{8I(2I-1)J(2J-1)} \nonumber \\
& &+ C_h^{173}\frac{5K^3+20K^2+4K(I(I+1)+J(J+1)+3-3I(I+1)J(J+1))-20I(I+1)J(J+1)}{4I(I-1)(2I-1)J(J-1)(2J-1)}
\end{eqnarray}
\end{widetext}
with $\nu_{\rm ave}$, $A_h^{173}$, $B_h^{173}$, and $C_h^{173}$ being constants that can be determined by measurements, $I$ being nuclear spin, $J$ being total electronic angular momentum, $K=F(F+1)-I(I+1)-J(J+1)$, and ${\bf F}={\bf I}+{\bf J}$. The second, third, and fourth terms correspond to the magnetic dipole, electric quadrupole, and magnetic octupole components, respectively. 

The formalism for the second-order hyperfine interaction is described in Ref. \cite{PhysRevA.77.012512}. Briefly, the shift is described as 
\begin{equation}\label{Eq2ndHF}
W_F^{(2)}=\sum_{\gamma'J'}\frac{|\langle \gamma' I' J' F M_F|H_{HFI}|\gamma IJFM_F \rangle|}{E_{\gamma J}-E_{\gamma'J'}},
\end{equation}
where $\gamma$ is the set of quantum numbers for electronic orbitals other than $I$, $J$, $F$, and $M_F$; $H_{\rm HFI}$ is the Hamiltonian for the hyperfine interaction; and $E_{\gamma J}$ is the energy for the state $\gamma J$. When the case of $J=2$ is considered, contributions from the adjacent fine structures dominate the sum over $\gamma'J'$. When we consider only the dipole-dipole interaction and dipole-quadrupole interactions, Eq. (\ref{Eq2ndHF}) results in 
\begin{align}
W_F^{(2)}=&
\left|
\begin{Bmatrix}
F 	& J	& I \\
1	& I	& J-1
\end{Bmatrix}
\right|^2\eta \nonumber \\
&+
\begin{Bmatrix}
F 	& J	& I \\
1	& I	& J-1
\end{Bmatrix}
\times
\begin{Bmatrix}
F 	& J	& I \\
2	& I	& J-1
\end{Bmatrix}
\zeta. \nonumber
\end{align}
$\eta$ and $\zeta$ are $F$-independent terms defined as 
\begin{align}
\eta =& \frac{(I+1)(2I+1)}{I}\mu^2 \frac{|\langle \gamma J-1 || T_{1}^e || \gamma J\rangle|^2}{E_{\gamma J}-E_{\gamma J-1}} \nonumber \\
\zeta=& \frac{(I+1)(2I+1)}{I}\sqrt{\frac{2I+3}{2I-1}} \nonumber \\
      &\times \mu Q \frac{\langle \gamma J-1 || T_{1}^e || \gamma J\rangle\langle \gamma J-1 || T_{2}^e || \gamma J\rangle}{E_{\gamma J}-E_{\gamma J-1}} \nonumber
\end{align}
with $T_1^e$ the spherical tensor of rank 1 acting on electronic coordinates. This leads to
\begin{align}
W_{F}^{(2)}=\left\{
\begin{array}{ll}
0 													& (F=9/2) \\
\frac{3}{140} \eta + \frac{3}{140\sqrt{2}} \zeta 	& (F=7/2) \\
\frac{32}{1575} \eta - \frac{8}{525\sqrt{2}} \zeta 	& (F=5/2)	\label{Eq2ndHFFit} \\
\frac{1}{100} \eta - \frac{1}{50\sqrt{2}} \zeta 	& (F=3/2) \\
0 													& (F=1/2) 
\end{array}
\right.
\end{align}
The contribution of the second-order hyperfine interaction to the A, B, C, and D constants can be calculated by first writing down $\nu_{F-1}-\nu_{F}$ ($F=3/2, 5/2, 7/2, 9/2$) as a function of hyperfine constants. Solving this for the hyperfine constants, the hyperfine constants are described as a function of $\nu_{F-1}-\nu_{F}$. Using the same inverse matrix on $W_{F-1}^{(2)}-W_{F}^{(2)}$, the following contributions of the second-order hyperfine interaction on the A, B, and C constants are obtained: 
\begin{align}
A:& \frac{1}{1050} \eta -\frac{2\sqrt{2}}{2625} \zeta \nonumber \\
B:& \frac{8}{315} \eta +\frac{\sqrt{2}}{105} \zeta \nonumber \\
C:& \frac{1}{350\sqrt{2}} \zeta \nonumber
\end{align}
In the analysis in the main text, $W_{F}^{(2)}$ described in Eq. (\ref{Eq2ndHFFit}) is added to the right-hand side of of Eq. (\ref{EqHFABC}). 


\bibliography{IS431}

\end{document}